\documentclass[a4paper]{spie}  %>>> use this instead for A4 paper
\usepackage{amsmath,amsfonts,amssymb}
\usepackage{graphicx}
\usepackage[colorlinks=true, allcolors=blue]{hyperref}

\def\CN2{\mbox{$C_N^2 $}}
\def\CT2{\mbox{$C_T^2$}}

\title{Optical turbulence forecast for ground-based astronomy and free-space optical communication}

\author[a]{Elena Masciadri}
\author[a]{Alessio Turchi}
\author[a]{Camilo Weinberger}
\author[a]{Marlene De Sepibus}
\author[a]{Luca Fini}
\affil[a]{INAF - Osservatorio Astrofisico di Arcetri, L.go E. Fermi 5, 50125 Firenze, Italy }

\authorinfo{Further author information: (Send correspondence to Elena Masciadri - E-mail: elena.masciadri@inaf.it, Telephone: +39 055 2752203}

\begin{document} 
\maketitle

\begin{abstract}
Forecasting optical turbulence in the Earth's atmosphere has been an ambitious challenge for the astronomical scientific community for several decades. While earlier research primarily focused on whether it was possible to predict optical turbulence and its vertical distribution, current efforts are more concentrated on the accuracy achievable at different timescales, the efficiency of various forecasting methods and the contributions of new statistical approaches, such as auto-regression and machine learning to this field. 
In this contribution, I will present the state of the art of the research conducted by our group, positioned within the international research scenery. Most of our past activity has been primarily focused on ground-based astronomy but recent advancements in space research opened new opportunities for applications in the free-space optical communication.
\end{abstract}


\section{INTRODUCTION}
\label{sec:intro}  % \label{} allows reference to this section

Over the last couple of decades most of major results and progress obtained in the field of the optical turbulence forecasting have been achieved in the context of the ground-based astronomy. In recent years, however, we have witnessed a significant increase in research activity in the field of the free-space optical communication (FOS) thanks, in part, to access to a huge number of satellites, particularly in Low Earth Orbit (LEO) satellites. It is therefore clear that in the coming years, the optical turbulence forecast will play a crucial role in FOS as the turbulence and the cloud cover remain the two main limitation for FOS.

Let's us begin by explaining why optical turbulence forecast plays a crucial role in ground-based astronomy. The main scientific drivers in the astronomical context can be summarised as follows: astronomers observational programs, particularly those aimed at the most challenging scientific goals, often require excellent turbulence conditions to be executed. 'Excellent conditions' refer to weak turbulence conditions. Unfortunately, the traditional queue system, which does not take into account turbulence conditions but only the scientific quality of programs, leads to a paradox that can be synthesised as follows: the most important is the challenge of a scientific program, the lower is the probability that the program can be carried out. This is the reason why it is a common conviction within the astronomical community that the Service Mode, which takes into account turbulence conditions in addition to the quality of scientific program, is essential to optimise the exploitation of top class telescopes of present time and Extremely Large Telescope (ELT) i.e. telescopes of new generation. There are, however, additional motivations among the scientific drivers for the OT forecasting. We know that the adaptive optics (AO) techniques can reach at present excellent performances (it is frequent a Strehl Ratio (SR) of 90\% in H band), however, the AO performances strongly depend on turbulence conditions. It is therefore crucial to be able to predict the state of the turbulence in advance if we want to fully exploit the adaptive optics potentialities. Finally, we highlight that the cost of a single night of observations for a top-class telescope of present time is of the order of one hundred thousand US dollars therefore an impressive amount of money. In synthesis: (1) we need to know optical turbulence conditions in advance to optimise observations and (2) all top-class telescopes are using or plan to use the Service Mode. That leads to the following conclusions: (a) the optical turbulence forecast is fundamental for the top-class telescopes and ELTS, (b) measurements do not access to the future as forecasts do and (c) non-hydrostatic models are the most suitable tool to achieve such a goal as they are based on the physics of the atmosphere. 

%%%%%%%%%%%%%%%%%%%%%%%%%%%%%%%%%%%%%%%%%%%%%%%%%%%%%%%

\section{Non-hydrostatic atmospheric meso-scale models}
\label{meso}

Which models are the most suitable for forecast the optical turbulence i.e. the $\CN2$ profiles ? We refer readers to a more detailed discussion presented in \cite{masciadri2013} that we summarize here. The most common atmospheric models are the General Circulation Models (GCMs) which are extended on the whole globe and are widely used to forecast classical atmospheric parameters. Their highest horizontal resolution is of the order of 10~km. Optical turbulence, however, is a random and strongly non linear phenomenon, characterised by spatial and temporal fluctuation scales that are much smaller than typical scales with which are reconstructed and resolved classical atmospheric parameters with GCMs. In spatial terms, the scales of optical turbulence can reach down to a few centimetres or even less and, in temporal terms, down to the millisecond. How can we bridge the gap between these two parameters space ? We can not apply on the whole world an atmospheric model with a resolution as high as the typical spatial scales of the optical turbulence fluctuation as this should require too large computing resources. The solution is, from one side, to use meso-scale atmospheric models therefore models applied to a limited surface area in a region around the point of interest on a surface of a few tens of kilometers with a sub-kilometric horizontal resolution (typically $\sim$ 500 m) and, on the other side, we parameterise the optical turbulence i.e. the $\CN2$. The parameterisation consists in expressing the fluctuations of microscopic quantities of the parameter in which we are interested on (for example the potential temperature as indicated in Eq.\ref{eq1}) as a function of the gradient of the same parameter but space averaged over a larger surface i.e. the model unit cell in our case:

\begin{equation}
w^{'}\theta^{'}=K\frac{\partial \theta}{\partial z}
\label{eq1}
\end{equation}

The factor K is a constant that characterises the turbulence properties of the turbulence of the specific fluid, in our case the atmosphere. Equation \ref{eq1} is known as {\it eddy diffusivity} approach. What we have just described is the basic idea of mesoscale atmospherical model. The necessity of the non-hydrostatic feature is due to the fact that hydrostatic models suppress vertical accelerations therefore suppress or deform gravity waves that are among the main sources triggering optical turbulence. 

Around 25 years ago it has been proposed the first optical turbulence scheme called Astro-Meso-Nh\cite{masciadri1999a,masciadri1999b} applied to a mesoscale atmospheric model called Meso-Nh\cite{lafore1998,lac2018} (that is a non-hydrostatic atmospheric model).  The optical turbulence scheme is based on a parameterisation that has been described in \cite{masciadri1999a,masciadri1999b} and upgraded on 2017 \cite{masciadri2017}. Astro-Meso-Nh model allows to calculate the volumetric distribution of the $\CN2$ in a volume around the point of interest and the associated integrated astroclimatic parameters\footnote{The term 'astroclimatic parameter' is commonly used in the astronomical community to distinguish the atmospheric parameters from those associated to the optical turbulence. The term might however be misleading as the optical turbulence is not univocally related to the ground-based astronomy context. We maintain the use of the term in the paper as it is commonly accepted in the astronomical community.}. 
All the astroclimatic parameters depend on the integral of the $\CN2$ along the whole atmosphere (around 20~km) weighted by a function F that can depend on the wind speed, or the height or the dynamical outer scale at different power laws as Eq.\ref{eq2}:

\begin{equation}
\int_{0}^{\infty} C_{N}^{2}(h) \times F\left(V^{a},h^{b},L^{c}  \right) dh
\label{eq2}
\end{equation}

A few among the most relevant astroclimatic parameters are: the seeing ($\varepsilon$) and the Fried's parameter ($r_{0}$) that from different perspectives represent the total amount of turbulence, the isoplanatic angle ($\theta_{0}$) that represent the angle within which the wavefront is coherent, the wavefront coherence time ($\tau_{0}$) that tells us how fast is the turbulence, the scintillation rate ($\sigma^{2}_{I}$) that quantifies the fluctuation of the intensity and some further parameters more specifically related to the ground-based astronomy such as the spatial coherence outer scale ($\mathcal{L}_{0}$)\cite{borgnino1990} and the isoplanatic angle for the multi-conjugated adaptive optics ($\theta_{M}$). In general the integral is calculated along the line of sight and then corrected by the air-mass factor to obtain estimates with respect to zenith.

We report below the main steps on which the Astro-Meso-Nh optical turbulence scheme is based on. We start from the prognostic turbulence kinetic energy Eq.\ref{eq3}:

\begin{equation}
\frac{\partial e}{\partial t}=-\overline{w^{'}u^{'}}
\frac{\partial U}{\partial z}-\overline{w^{'}v^{'}}
\frac{\partial V}{\partial z}+\frac{1}{\rho}\frac{\partial(\rho w^{'}e^{'}) }{\partial z}-0.7\frac{e^{3/2}}{L}+\frac{g}{\theta_{v}}\overline{w^{'}\theta_{v}^{'}}
\label{eq3}
\end{equation}

done by the shear production factor (first two terms on the right side), diffusion factor (third term), viscous dissipation factor (fourth term) and buoyancy energy factor (fifth term). The buoyancy factor is the most relevant for the optical turbulence that is indeed triggered by buoyancy forces. 

By applying the eddy diffusivity approach it has been possible to prove that the $\CT2$ is equal to Eq.\ref{eq4}\cite{masciadri1999a,masciadri1999b,masciadri2017}:

\begin{equation}
C_{T}^{2}(h)=0.58\phi_{3}L^{4/3}(h)(\frac{\partial\theta (h)}{\partial h})^{2}
\label{eq4}
\end{equation}

where L is the mixing length (in this case the BL89 mixing length\cite{bougeault1989}), $\theta$ the potential temperature and $\phi_{3}$ is thermo-stability term proportional to the inverse of the Prandtl number. Finally, using the Gladstone's law, we obtain the $\CN2$ (Eq.5)\cite{masciadri2017}:

\begin{equation}
C_{N}^{2}(h)=\left( \frac{80\times 10^{-6}P(h)}{T(h)\theta(h)} \right)^{2}C_{T}^{2}(h)
\label{eq5}
\end{equation}

Such an approach to forecast the optical turbulence based on the computation of the $\CN2$ starting from the prognostic turbulence kinetic energy (TKE) and a mixing length (that works for stable and unstable regimes) and a spatio-temporal evolution of an hydrodinamical equation of the atmospheric flow on the whole 20~km has been later on used by other teams/studies\cite{cherubini2008,basu2020}. This has been done with different atmospherical models (WRF in the specific case) therefore also different mixing lengths. We are however moving on the same typology of approach.
Other methods to forecast the optical turbulence have been used in the literature and in the astronomical context to reconstruct the $\CN2$ over the whole atmosphere but they are mainly analytical equation depending on atmospheric parameters that are supported by an empirical fit with measurements\cite{trinquet2007,giordano2013,osborn2018} and they do not involve a turbulence scheme in a hydrodinamical code.

\subsection{Model Calibration and model validation}

On 2001, for the first time it has been presented the thesis of the necessity of a calibration for a non-hydrostatic atmospheric model to reconstruct reliable $\CN2$ estimates on the whole around 20 km above the ground\cite{masciadri2001}. We developed during the years different strategies for the model calibration that are, however, based on the same principle exposed in Masciadri \& Jabouille 2001\cite{masciadri2001} that states that in the free atmosphere we need to provide to the model the structure of the $\CN2$ at large spatial scales. It is as if the model needs to know the skeleton of the $\CN2$ over which it can develop during each simulation the optical turbulence of the specific night. The model calibration can be performed with $\CN2$ measurements and it is based on the fact that Masciadri \& Jabuille (2001)\cite{masciadri2001} proved that those parts of $\CN2$ that are in stable regime are proportional to TKE$_{min}^{2/3}$ (where TKE$_{min}$ is a seed used to trigger the turbulence kinetic energy scheme). Measurements of the $\CN2$ related to a rich statistical samples of several ten of nights derived from vertical profilers such as Generalized SCIDAR - GS\cite{avila1997} (or more recently also with Stereo-SCIDAR\cite{sheperd2014}) are used to calibrate models. Once the calibration has been performed, the free parameters of the model are fixed and models have to be validated. To validate a model it is necessary to compare forecasts and observations on a rich statistical sample of independent nights. The means that the sample of nights has to be independent from the sample with which the model has been calibrated. In the last 20 years significant progress have been done and it is now possible to collect samples of measurements extended on a several ten of nights up to hundreds of nights\cite{garcialorenzo2011a,garcialorenzo2011b,osborn2018b}. 
In the following studies\cite{masciadri2004,hagelin2011,masciadri2017} our team performed model validation using some slightly different calibration strategies above different sites respectively at San Pedro Martir (Baja California), Mt.Graham (Arizona) where is the Large Binocular Telescope (LBT) and Cerro Paranal (Chile) where is the site of the Very Large Telescope (VLT). It is worth to highlight that calibration performed with richer statistical sample of measurements are more representative and efficient. 

%\subsection{Most relevant achievements}

%%%%%%%%%%%%%%%%%%%%%%%%%%%%%%%%%%%%%%%%%%%%%%%%%%%
\section{Automatic operational forecasts systems and hybrid methods}
\label{hybrid}

Between the end of last century and beginning of the current one the typical open questions was wether if we were able to forecast reliable $\CN2$. In recent years, however, the typical open questions are more addressed on our ability in implementing automatic forecast systems of the OT and on the ability in achieving a required accuracy/performance. 

\subsection{ALTA Center at the Large Binocular Telescope}

\begin{figure*}
\begin{center}
\includegraphics[angle=-90,width=0.6\textwidth]{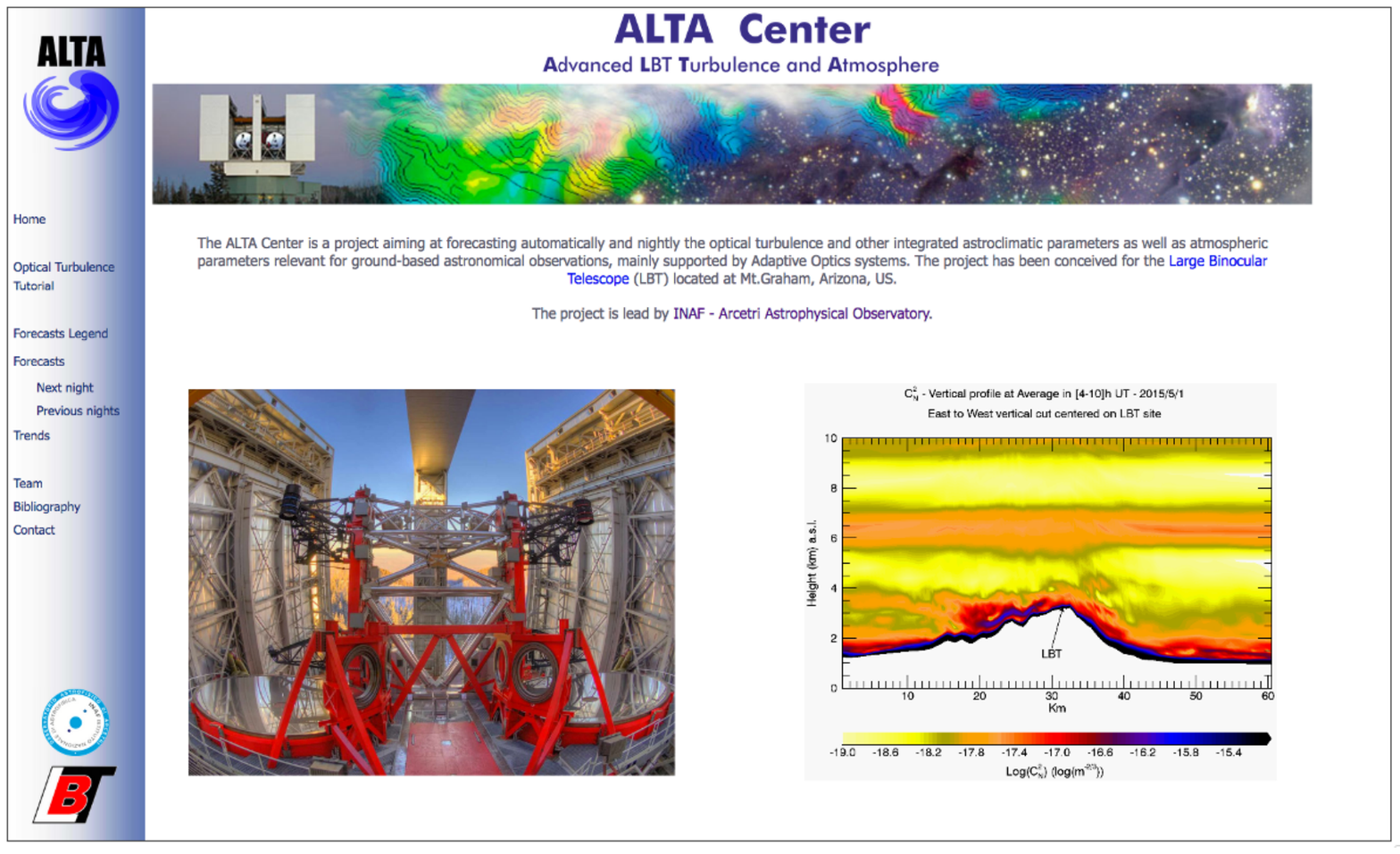}\\
\end{center}
\caption{ALTA Center web page home: \href{http://alta.arcetri.inaf.it}{http://alta.arcetri.inaf.it}.}
\label{alta}
\end{figure*}

\begin{figure*}
\begin{center}
\includegraphics[angle=-90,width=\textwidth]{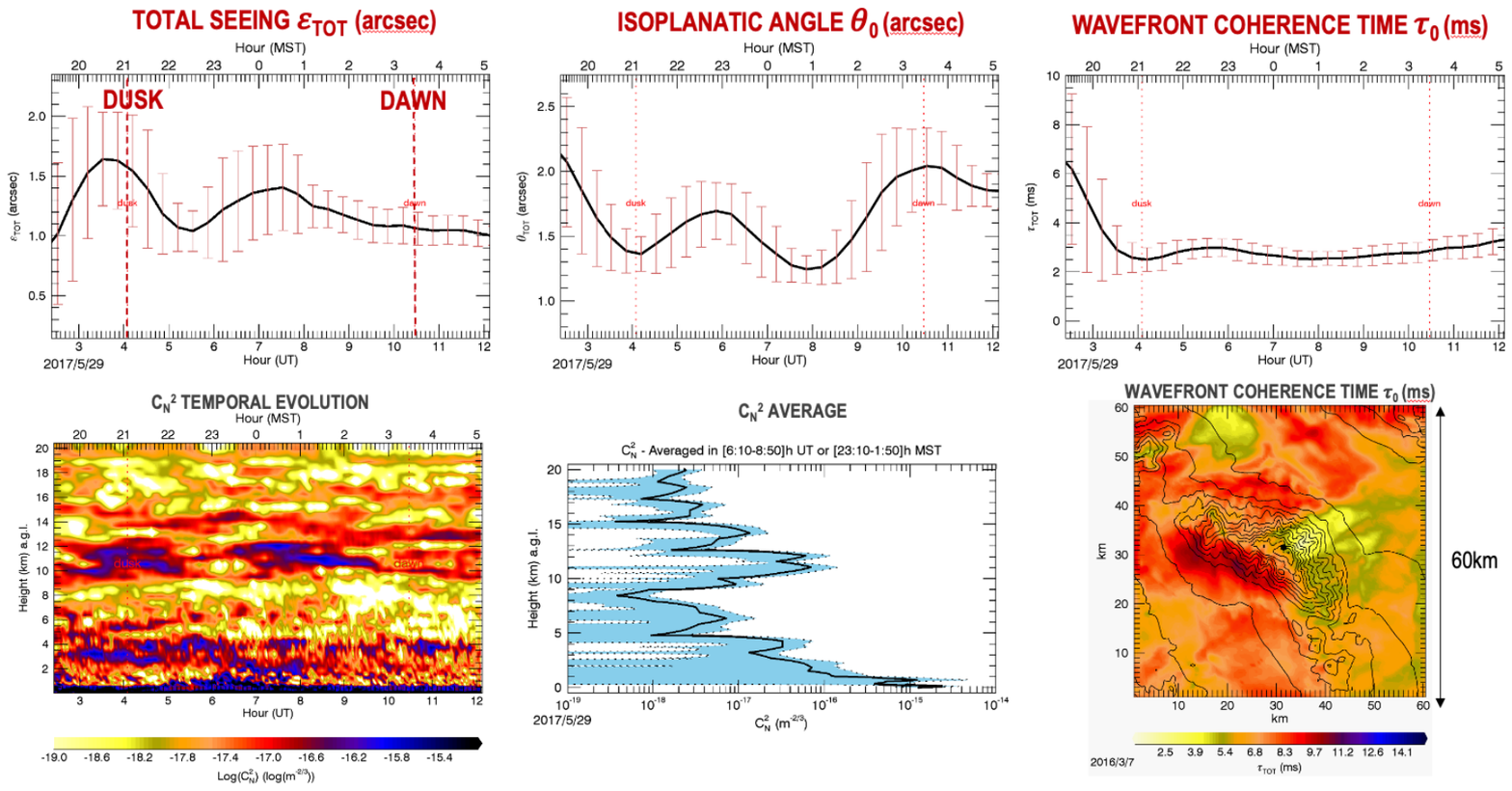}\\
\end{center}
\caption{Examples of outputs/products provided by ALTA Center - Top: temporal evolution between the dusk and dawn of the total seeing ($\varepsilon$), isoplanatic angle ($\theta_{0}$) and wavefront coherence time ($\tau_{0}$). Bottom: temporal evolution of the $\CN2$ on the 20~km above the ground (left), average of the $\CN2$ profile all along a night (centre), average all along one night of the wavefront coherence time extended on a surface of 60 km $\times$ 60 km (right).}
\label{examples_alta}
\end{figure*}

ALTA Center has been conceived to support the science operation of the LBT (Arizona, US) as stated by the Director of the LBT\cite{veillet2016}. Development of ALTA lasted about one year and the system is operational since 2016 i.e. almost 10 years. It is a fully automatic operational forecast system for the OT and all atmospheric parameters relevant for the ground-based astronomy running nightly. Figure \ref{alta} shows a snapshot of the of the website. The main information available to people of the LBT consortium inside ALTA Center includes the forecasts of the astroclimatic parameters and atmospheric parameters of the coming night as well as forecasts related to a few nights in the past. Fig.\ref{examples_alta} shows an example of a few outputs/products provided by ALTA Center. On top (from left to right) is shown the temporal evolution between dusk and dawn of the total seeing, the isoplanatic angle and the wavefront coherence time. On the bottom (from left to right) is shown the temporal evolution of the $\CN2$ during the night extended on the whole $\sim$ 20~km, the average of the $\CN2$ profile during the night with its variability and a 2D map extended on a surface of 60~km $\times$ 60~km of the wavefront coherence time ($\tau_{0}$). A validation of ALTA Center for the atmospheric parameters has been published at the epoch of the commissioning of the system\cite{turchi2017}. The calibration and validation of the model has been carried out using measurements done with a Generalised SCIDAR related to an extensive site testing campaign\cite{masciadri2010}. 
So far we are considering forecasts of the OT  that cover the entire night. More precisely, at a time T$_{disp}$, prior to the beginning of the night, we deliver the forecast of the OT i.e. the temporal evolution of the forecast through the night. It is however necessary, at this point, to explain which is the role of {\it 'time'} in a forecast process. 

\subsection{The time scale in a forecast process}
\label{fts_sec}

\begin{figure*}
\begin{center}
\includegraphics[angle=90,width=0.8\textwidth]{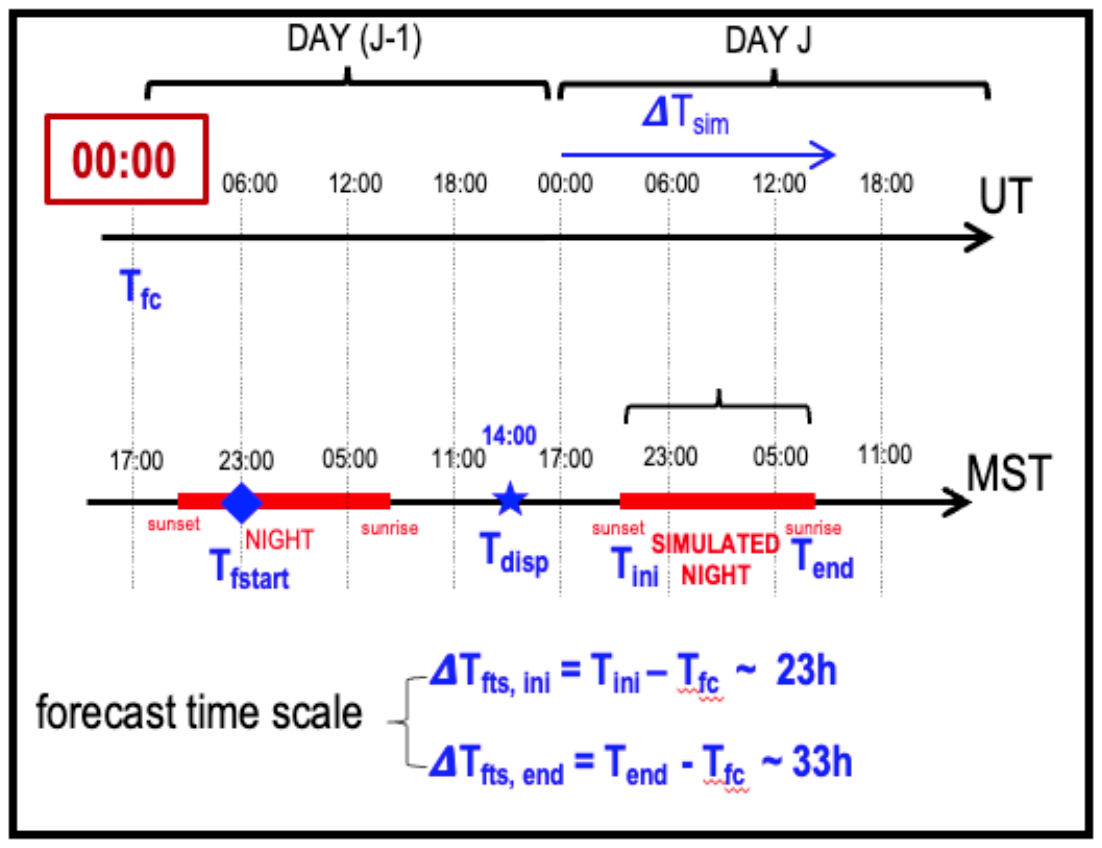}\\
\end{center}
\caption{Extracted from Masciadri et al. 2023\cite{masciadri2023}. Toy model: model configuration related to the automatic forecast system of the Very Large Telescope (FATE project). See text in Section \ref{fts_sec}.}
\label{fts}
\end{figure*}

In the perspective to quantify the accuracy of an optical turbulence forecast system it is crucial to introduce the concept of Forecast Time Scale (FTS) as this is strictly related to our ability to assess forecast performances. Without a clear definition of FTS any attempt to quantify accuracy would be meaningless. Several different types of time and time intervals play a role in the forecasting process. In the toy model illustrated in Fig.\ref{fts} are reported the main ones. Let us assume a temporal axis which can be expressed either in UT or LT. We can distinguish: (1) the time in which the initialisation data of the numerical atmospherical model are calculated, (2) the time in which the simulation starts, (3) the simulated time that is an interval of time that is reconstructed, (4) the time required to calculate the forecast i.e. the effective computation time, (5) the time in which a forecast is displayed i.e. the time when the forecast becomes available to the user and finally, (6) the forecast time scale (FTS) i.e. the interval of time between the time in which the initialisation data are calculated and the time the forecast refers. In Fig.\ref{fts} is sketched the case of one of our studies \cite{masciadri2023} in which it is described the configuration of the model used for the automatic forecast system for the Very Large Telescope (FATE project - see Section \ref{fate}). Let us assume we are in the date J-1 and that we want to forecast the optical turbulence during the coming night bounded by T$_{ini}$ and T$_{end}$. In this configuration the forecast time scale ranges between 23h and 33h. We refer the reader to Masciadri et al. 2023\cite{masciadri2023} for an extensive description of the concept of the FTS. Here we emphasize that the FTS depends on many factors, including the longitude of the site, the model configuration and the products to be delivered that determine the sequence of forecasts to be executed. We highlight that, for example, in a different automatic forecast system that we implemented at the Large Binocular Telescope (LBT) in Arizona (ALTA Center) the FTS$_{LBT}$ = FTS$_{VLT}$ - 9h. That means that in any case the FTS is typically of an order of more than 10h. The key question is, therefore, wheter forecasts can be performed at shorter forecast time scale FTS but with improved accuracy. This issue is particularly timely since timescales of 1-2 hours are the most relevant for the science operation of current top-class telescopes and future ELTs.

\begin{figure*}
\begin{center}
\includegraphics[width=0.5\textwidth]{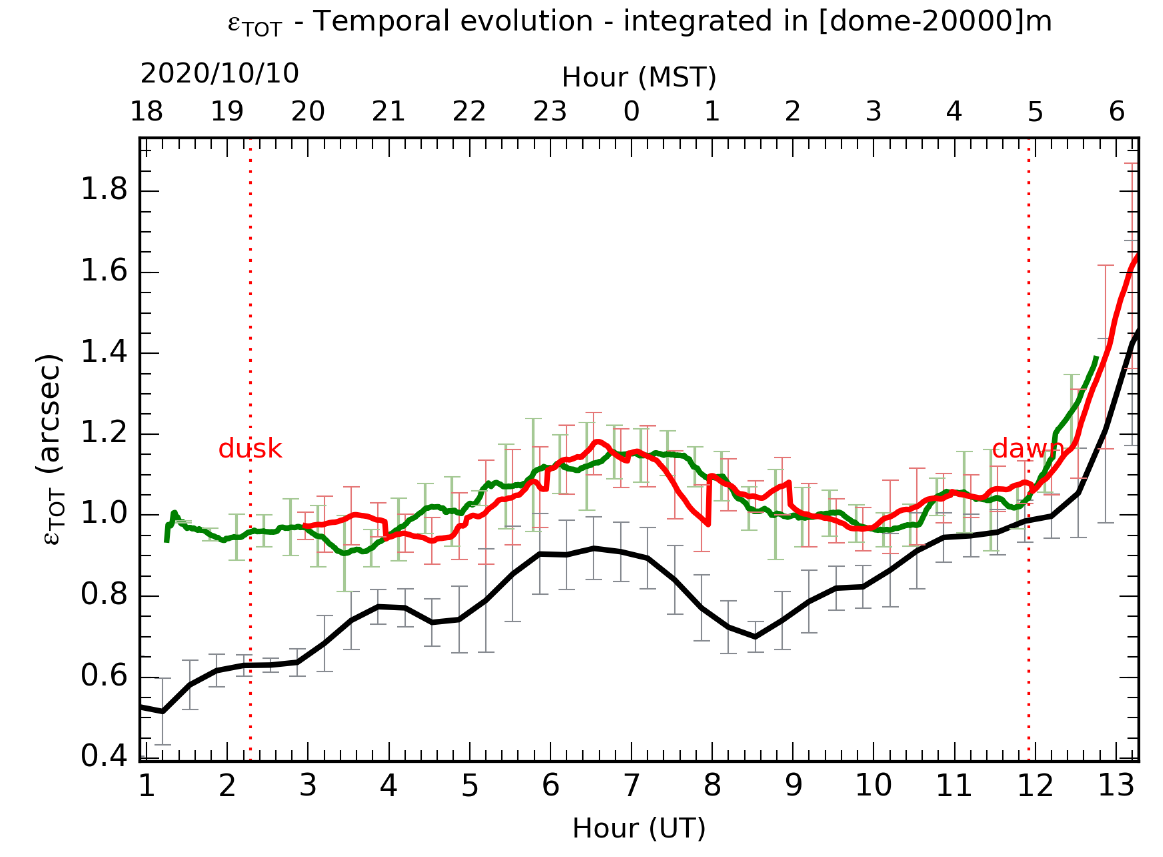}\\
\end{center}
\caption{Temporal evolution of the seeing during one night: green line: real-time observations; black line: forecast of the seeing obtained with the atmospheric Astro-Meso-Nh model in standard configuration therefore the afternoon for the coming night; red line: forecast of the seeing obtained with the AR method calculated at a time scale of 1 hour. }
\label{ar_seeing}
\end{figure*}

%%%%%%%%%%%%%%%%%%%%%%%%%%%%%%%%%%%%%%%%%%%%%%%%%%%%%

\subsection{The AR method}

In this framework we proposed in the last years a new technique based on an auto-regressive filter technique Masciadri et al. 2020\cite{masciadri2020} that we called AR. Following this approach the forecast with the AR method at the time (t+1) is equal to the forecast provided by the atmospherical model Astro-Meso-Nh model at the time (t+1) plus a factor that depends on the difference between the Astro-Meso-Nh atmospheric model prediction and observations through a number of coefficients called regressors. With observations we mean real-time observations related to a finite number of nights in the past. With this in mind the new operational forecast system consists on the following: in the afternoon we can provide the forecast for the coming night as usual (that means the temporal evolution on the whole night). We call that forecast 'forecast in standard configuration'. Then, when the night starts, i.e. when real-time measurements start to be available, at each full hour the forecast calculated with the AR technique extended on the successive 4h is up-graded. If we repeat the up-graded forecast with the AR technique at each full hour we obtain a forecast at a time scale of 1h. In Fig.\ref{ar_seeing} is shown an example obtained with the case of the seeing during one night above the VLT. The black line represent the forecast obtained in standard configuration therefore the forecast of the seeing related to the coming night J (the night starts on the date (J-1) and ends on the date J) calculated in the afternoon of the date (J-1). The green line represent the real-time measurements obtained with the Differential Image Motion Monitor (DIMM) and the red line represents the forecast obtained with the AR technique. It is possible to appreciate the clear improvement provided by the AR technique. In the study applied to the LBT site by Masciadri et al. 2020\cite{masciadri2020} it has been possible to prove that the gain of the AR technique on a time scale of 1h with respect to the standard configuration is a factor between 2 and 8 depending on the parameter we are interested on.  The method proposed has been extensively validated and it is has been proved that we obtain the maximum gain if we use observations from a number N = 5 night in the past in the calculation of the AR\cite{masciadri2020}. Such a system has been implemented in the operational forecast system ALTA Center and since 2019 (around 6 years) ALTA Center provides forecasts at different time scales: at short and long time scales:\\ \\
\noindent
{\bf - Long time scale}: at around 14:00 LT forecasts of all the astroclimatic and atmospheric parameters related to the whole coming night are displayed.\\ 
\noindent
{\bf - Short time scale}: forecast of the same parameters extended on the successive 4h are displayed and up-graded hourly. Since August 2024 we replaced the AR forecast up-grade having a frequency of 1h with a frequency of 10 minutes. \\ \\
\noindent
At our knowledge, ALTA Center is the first site providing automatic operational forecasts of the optical turbulence to support the science operation of top-class telescopes at long and short forecast time scales as described with similar performances. It is of course obvious that the AR technique can be applied to all parameters for which we can access to real-time measurements done above a specific site. Above the site of LBT (Mt.Graham) there are continuous real-time measurements for temperature, wind speed and direction, relative humidity and for the seeing that is, however, the unique astroclimatic parameter. We have not real-time measurements of the precipitable water vapour (PWV) neither the other astroclimatic parameters for which is necessary a vertical profiler. Luckily, our team developed more recently a different and new automatic operational forecast system for the Very Large Telescope (at Cerro Paranal, Chile) that is called FATE project that could provide answer to these limitaitons. 

\begin{figure*}
\begin{center}
\includegraphics[width=0.6\textwidth]{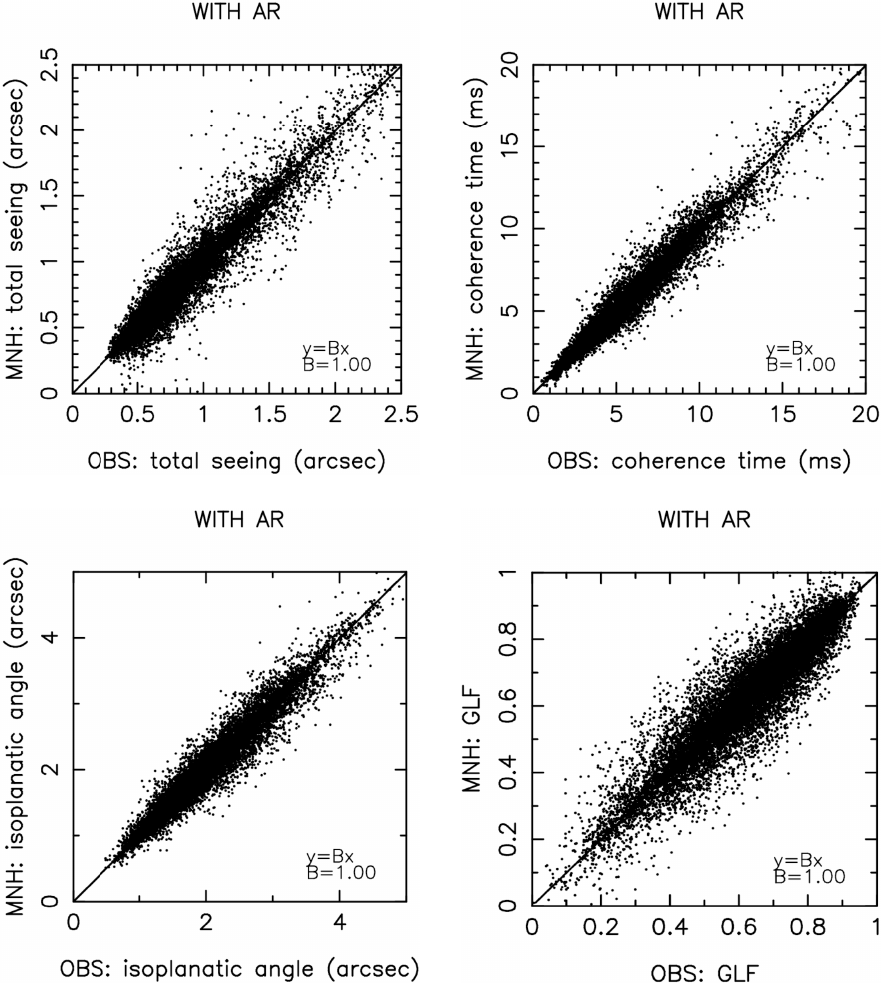}\\
\includegraphics[width=0.7\textwidth]{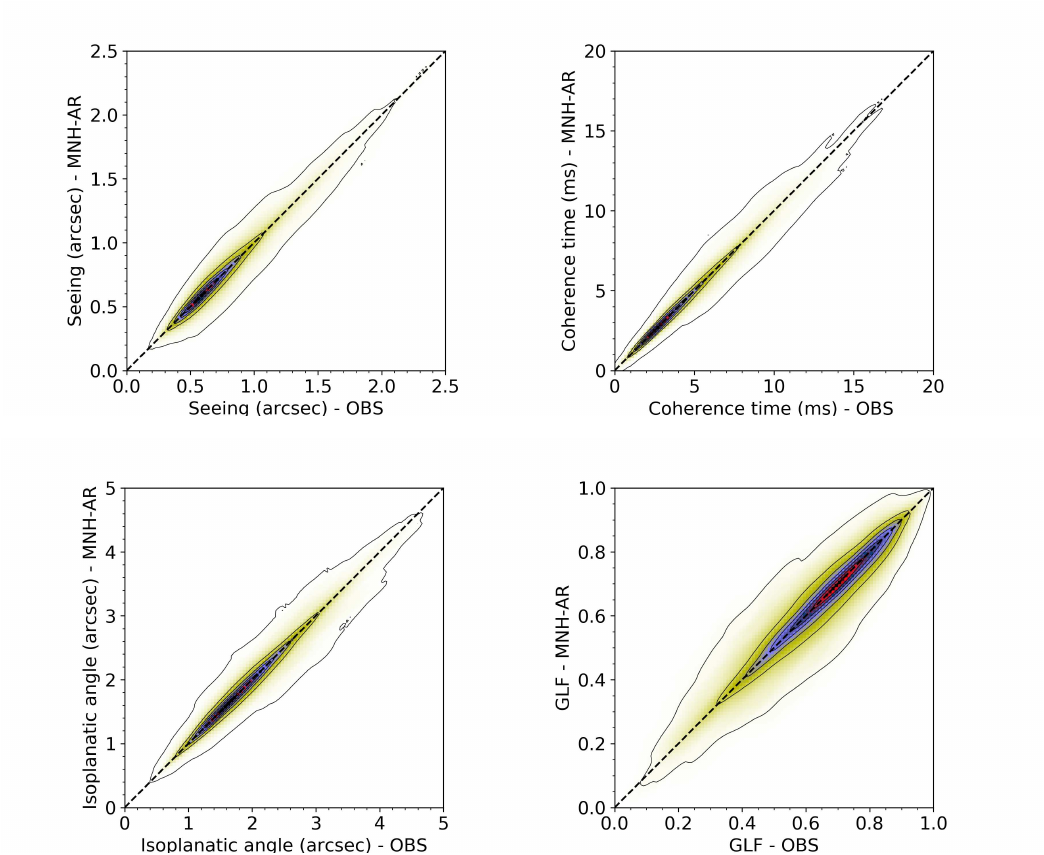}\\
\end{center}
\caption{Extracted from Masciadri et al. 2023\cite{masciadri2023}: Top: scatters plots of forecast versus observations for seeing, wavefront coherence time, isoplanatic angle and GLF. Bottom: Density function maps for the same parameters shown on top.}
\label{ar}
\end{figure*}

\subsection{FATE at the Very Large Telescope} 
\label{fate}

The project FATE\cite{masciadri2024_fate}  won an international call opened by the European Southern Observatory (ESO). It is operative since June 2024 and it implies a continuous R\&D activity aiming to improve forecasts performances and/or adding delivery products according to the VLT necessities. The great added value of FATE is that the VLT is equipped with a vertical profiler (the Multi Aperture Scintillation Sensor - MASS) that, combined with a DIMM, can provide continuous real-time measurements of all the astroclimatic parameters i.e. the seeing ($\varepsilon$), the isoplanatic angle ($\theta_{0}$), the wavefront coherence time ($\tau_{0}$) and the ground later fraction (GLF)\footnote{Isoplanatic angle is not included because it has not been requested by ESO.}. The latter is the ratio between the turbulence developed close to the ground (J$_{G}$) versus the total turbulence developed all along the whole 20~km i.e. in the ground (J$_{G}$) and in the free atmosphere (J$_{F}$):

\begin{equation}
GLF = \frac{J_{G}}{J_{G}+J_{F}}
\end{equation}

where J is the integral of the $\CN2$ in a vertical slab. The GLF is relevant for all typologies of wide-field adaptive optics i.e. the Ground Layer Adaptive Optics (GLAO), the Laser Tomographic Adaptive Optics (LTAO), the Multi-Conjugated Adaptive Optics (MCAO), the Multi-Object Adaptive Optics (MOAO). This allowed us to implement the AR method to basically all the astroclimatic parameters plus the PWV.

Figure \ref{ar} show the quantification of the accuracy of the AR method for all the astro-climatic parameters calculated on a sample of one full solar year [2018/08/01, 2019/07/31] related to the nighttime\cite{masciadri2023} for a forecast time scale FTS = 1h. On top we have the scatter plot and on bottom the density function maps where it is visible that the peak of the distribution is well centered with respect to the bisector. That tells us that the are no biases. In Table \ref{table_ar} are shown the quantitative estimates of bias, RMSE and SD associated to Fig.\ref{ar}. We have a RMSE  = 0.1'' for the seeing, 0.58 ms for $\tau_{0}$, 0.16'' for $\theta_{0}$ and 6\% for the GLF. The analysis can be performed also in alternative way. It is possible to calculate the RMSE on each individual night, calculate the cumulative distribution on the sample of 1 solar year for all the nights and we finally retrieve the median and first and third quartiles of the RMSE for whatever time scale. This last approach provides better performances (i.e. smaller values of RMSE) but this is a just a different way to analyse data. The important thing is to declare which system is used. We just refer reader to the paper Masciadri et al. 2023\cite{masciadri2023} for supplementary infos on this analysis. The important point (and this is what we want to highlight here) is to quantify how good/bad are performances of AR forecasts i.e. how to quantify the quality of these forecasts. The unique serious way to do that is to compare uncertainty between forecasts and observations with the accuracy with which we can perform such a measurements therefore the uncertainty between measurements of the astroclimatic parameter provided by different instruments.

We could proved that on at the VLT, by taking a sample of 157 nights in which we have measurements of different astroclimatic parameters taken at the same time with a MASS-DIMM and with a Stereo SCIDAR (GS) and showing that rhe RMSE obtained between forecasts and observations is not larger than the uncertainty between the RMSE between the two instruments.  Table \ref{instr} reports the bias, RMSE and SD of the case we have just described. We can conclude that the SD of Table \ref{table_ar} is smaller than SD$_{obs}$ of Table \ref{instr} i.e. SD $<$ SD$_{obs}$.  We used SD instead of RMSE as we wanted to be conservative. Taking the SD instead of the RMSE we are basically assuming that all systematic uncertainty are equal to zero and we are treating only statistical uncertainties. That tells us that performances of forecasts are sufficiently good to provide an effective a useful information on a time scale of 1h. The conclusion of our study prove that the AR method can provide very accurate forecasts.

\begin{table}
\begin{center}
\caption{ \label{table_ar}Extracted from \cite{masciadri2023} - 
Statistical operators (BIAS, RMSE, SD) obtained with observations and the AR method on a time scale of 1h. Selection of the seeing $\le$ 1.5 arcsec is done on observations.
}
\begin{tabular}{cccccc}
\hline
      & $\varepsilon$  & $\varepsilon$  $\le$ 1.5 arcsec  & $\tau_{0}$ & $\theta_{0}$  & GLF   \\
       &   (arcsec) & (arcsec) & (ms) & (arcsec) & $\times$ 100 (\%)\\
\hline
BIAS & 0.00 & 0.00 & -0.01 & -0.01&  0.00 \\
RMSE  &  0.10&   0.09&   0.58 &  0.16& 0.06\\
SD  &      0.10 &   0.09 &   0.58&  0.16& 0.06\\
\hline
\end{tabular}
\end{center}
%\label{table_ar}
\end{table}

\begin{table*}
\begin{center}
\caption{ Extracted from \cite{masciadri2023} - Accuracy obtained with observations: Stereo-SCIDAR (SS) and MASS-DIMM. First row: BIAS$_{obs}$ calculated by comparing simultaneous values from different instruments calculated on a rich statistical sample. Second row: as the first row but it is shown the RMSE$_{obs}$. Third row: as the first row but it is shown the standard deviation SD$_{obs}$. }
\begin{tabular}{lcccc}
\hline
Param. & Seeing ($\varepsilon$)  &  Wavefront coherence time ($\tau_{0}$) & Isoplanatic angle ($\theta_{0}$)  & GLF   \\
   &   (arcsec) &  (ms) & (arcsec) & (x 100 \%)\\
\hline
Instruments   & SS vs. DIMM &   SS vs. MASS-DIMM & SS vs. MASS-DIMM & SS vs. MASS-DIMM \\
N. nights & 157    &  142 & 142 & 142 \\
\hline
BIAS$_{obs}$ &  0.08  &  0.41 & - 0.16 &  0.15 \\
RMSE$_{obs}$ &  0.25 &   1.29 &  0.45  & 0.20 \\
SD$_{obs}$ &  0.24 & 1.22  &  0.42  & 0.14 \\
\hline
\end{tabular}
\end{center}
\label{instr}
\end{table*}

%%%%%%%%%%%%%%%%%%%%%%%%%%%%%%%%%%%%%%%%%%%%%%%%%%%%%%%%

\subsection{Machine Learning}

Machine learning has been tested to evaluate wheter it might provide better results than the AR method. There is no space here to enter in details on this topic. However, it has been shown that the Random Forest algorithm not only performs worse than the AR method also does not outperform the reference method of the prediction by persistence above at the VLT (FATE project)\cite{masciadri2023,turchi2022}. A more extensive study is currently underway\cite{weinberger2025} in our team aiming to investigate alternative machine learning algorithms across different sites and this will be able to provide a more complete and accurate picture of the state of the art. We can conclude however that, on the base of results achieved so far, atmospheric models are still competitive and are not outperformed by machine learning. At the same time it is also worth to highlight that, independently of forecast performances, machine learning has the characteristics to be very flexible because of its nature and it appears interesting for some applications for which the use of other approaches might be difficult or impossible. It seems therefore always useful to invest on machine learning. The lesson learned so far is that it appears clear that we are entering a new era that will be dominated by hybrid techniques done by numerical technique i.e. based on hydrodynamic atmospheric models plus statistical techniques that can improve performances of the former on a shorter time scales. We include in the statistical category the AR method, the machine learning, deep learning, neural networks, etc.. At present the AR method presents the best performances therefore is our benchmark of reference.

\begin{figure*}
\begin{center}
\includegraphics[angle=-90,width=0.8\textwidth]{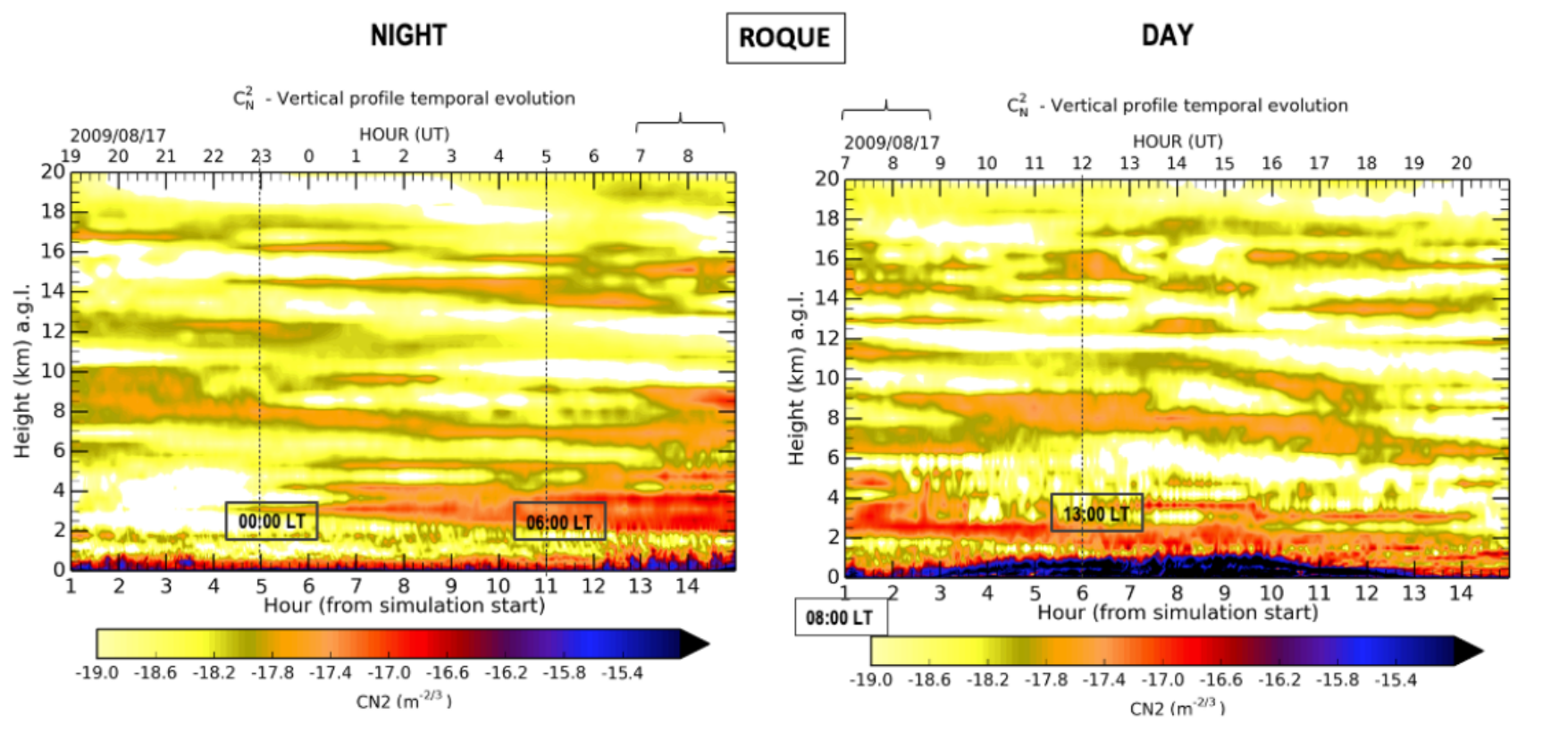}\\
\end{center}
\caption{Extracted from \cite{masciadri2024}. Top: Temporal evolution of the $C_N^2$ profiles above the Roque de los Muchachos Observatory (ORM) related to the 2009/08/17 date during the nighttime (left) and during the daytime (right). }
\label{cn2_roque}
\end{figure*}

\begin{figure*}
\begin{center}
\includegraphics[angle=-90,width=0.7\textwidth]{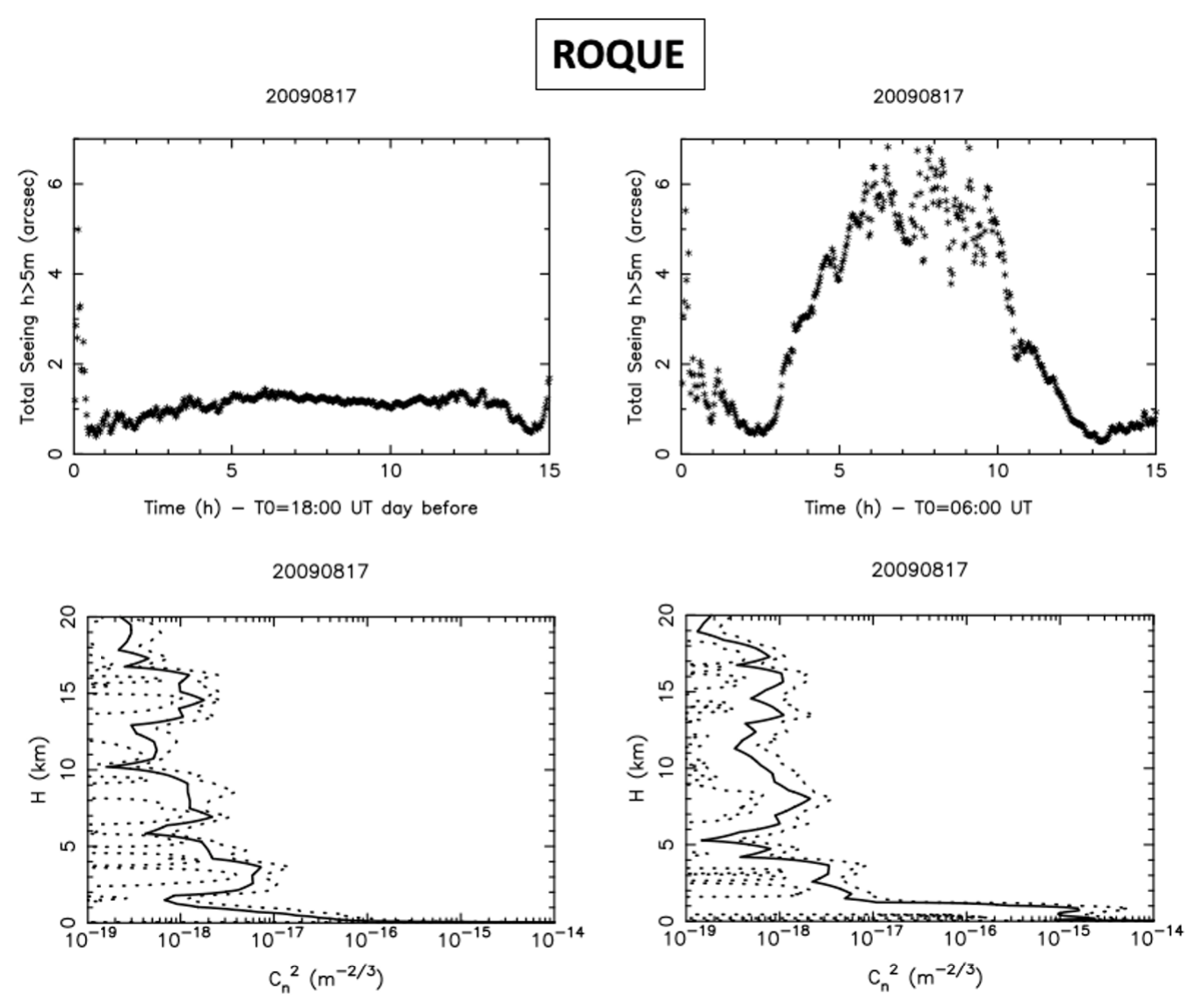}\\
\end{center}
\caption{Extracted from \cite{masciadri2024}. Top: Temporal evolution of the seeing related to the date 2009/08/17 above the Roque de los Muchachos Observatory (ORM) during the nighttime (left) and during the daytime (right). Bottom: average of the $C_N^2$ during the night (left) and during the day (right) of the same date. Dashed lines represent the standard deviation.}
\label{see_roque}
\end{figure*}

\begin{figure*}
\begin{center}
\includegraphics[angle=-90,width=0.7\textwidth]{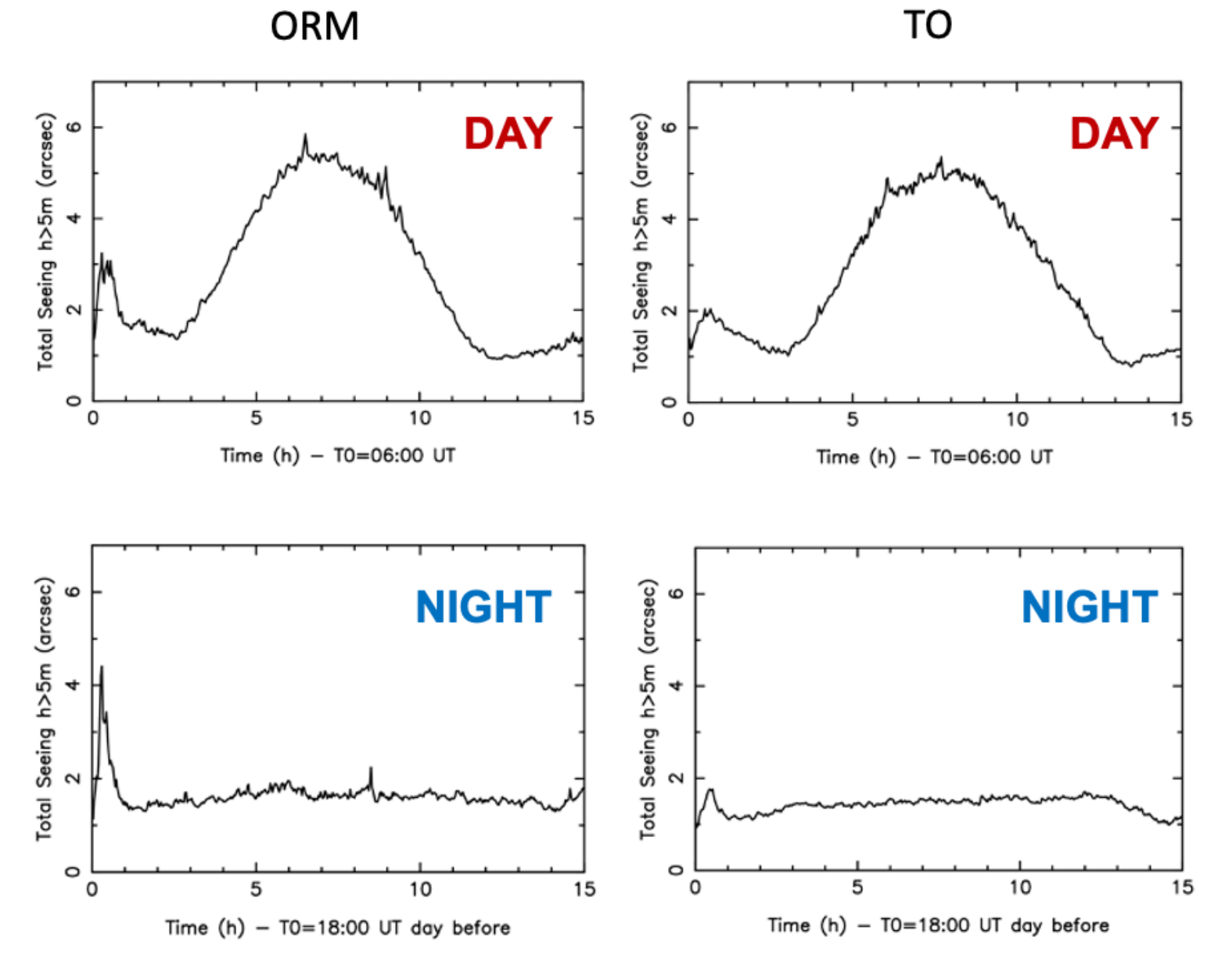}\\
\end{center}
\caption{Extracted from \cite{masciadri2024}. Top: Temporal evolution of the seeing during the daytime averaged on the same 20 dates above ORM (left) and TO (right). Bottom: Temporal evolution of the seeing during the nighttime averaged on the same 20 dates related to the daytime above ORM (left) and TO (right).}
\label{see_20_comp}
\end{figure*}

%%%%%%%%%%%%%%%%%%%%%%%%%%%%%%%%%%%%%%%%%%%%%%%%%%%%%

\section{Free-space optical communication (FOS): challenges for OT forecasts}
\label{fos}

Research activity related to free-space optical communication (FOS) is living in the last years an incredible boost thanks in part to the huge number of satellites (particularly Low Earth Orbit - LEO satellites) we can access to. FOS is particularly attractive with respect to the traditional communication in the radio-frequency (RF) regime allowing a few main advantages: (1) 
the bandwidth of transmission is much larger in the optical than in the RF regime (order of MHz or a few GHz in the RF while a few hundred of THz in the optical). That makes the amount of transmitted information about 10$^{3}$ greater in the optical than in the RF regime, (2) the transmission in the optical regime is also more secure thanks to a narrow beam propagation which makes interception of signal more difficult, (3) a reduced satellite payload weight allows for lower lunch and operational costs. 
If optical frequencies offer some advantages with respect to RF however turbulence and cloud cover represent the main sources of noise at these frequencies and for this reason the optical turbulence forecast is crucial to exploit space communication potentialities at these frequencies. It is indeed important that we determine the optimal time and geographical windows for satellite communication. We expect that the OT forecast will play a fundamental role in FOS in the coming years.

Our expertise set-up in the last decade in the astronomical field is certainly useful and can be transferred in this field. There are, however, challenges related to features specifically related to the OT forecast in FOS that do not concern the astronomical context. That means that when we study the OT forecast in FOS we have to tackle specific aspects that are negligible or irrelevant in the astronomical context. The main differences are listed below:
\begin{itemize}
\item We are supposed to forecast optical turbulence on sites characterised by whatever optical turbulence and weather conditions. This represents an important difference with respect to the astronomical case in which sites are, by definition, characterised by very favourable (i.e.weak) turbulence conditions, they are in general very dry sites with good photometric conditions that means minimum cloud cover. 

\item We are interested in forecasting optical turbulence on the entire 24-hours period and not only on the nighttime (as is typical in stellar astronomy). OT conditions during the day are very different from those at nighttime case, particularly close to the ground due to the solar heating during the daytime. During the day the atmosphere is predominantly in convective regime whereas during the nighttime it is mainly in stable regime. It is important that non-hydrostatic atmospheric mesoscale models respond properly at nighttime as well as at daytime. Also the OT parameterization might required modification in day time and  in any case it should be validated in this regime. 

\item While in the ground-based astronomy case we have a propagation of the wavefront within about $\pm$ 50 degrees with respect to zenith, in the FOS case wavefront propagation can be at angles as small as 5 or 10 degrees with respect to the horizon. While in the first case it is possible to apply approximation to the theory of the wavefront propagation through a turbulent medium that strongly simplify equations, in the second case this is not necessarily feasible and we might be in conditions in which the theory is much more complex. A classical example is the weak perturbations approximation that is valid in basically all applications in ground-based astronomy. When the wavefront propagates at very small angles with respect to the horizon this approximation might not be respected any more. We might also enter in a regime of strong saturation for the scintillation as an almost horizontal propagation through turbulence close to the ground (where we have most of the turbulence) makes this possible.

\item While in the ground-based astronomical context the wavefront propagation is only downwards, in FOS the wavefront propagation has a double direction i.e. downwards and upwards. Due to the fact that the optical turbulence is not uniformly distributed in the atmosphere (i.e. $\sim$ 20~ km) and there are almost two order of magnitude of difference between the $\CN2$ in the free atmosphere and in the boundary layer, the effects of the optical turbulence on the astroclimatic parameters such as (isoplanatic angle, wavefront coherence time, scintillation rate, etc.) is not symmetric with respect to the wavefront propagation (downwards or upwards). We can have, for example, a consistent scintillation rate in the upwards direction while the same parameter remains weak in the downwards direction as it is driven by the turbulence at high altitude. This is due to the fact that the scintillation depends on the distribution of the turbulence and it is particularly sensitive to turbulence close to the source and most part of the turbulence is concentrated close to the ground.
\end{itemize}

In order to tackle the issue related to the OT forecast extended on the 24-hours period we report results we obtained in the context of the EU SOLARNET project in Horizon 2020 framework. Our team performed a study on Canary Islands, more precisely above the Roque de los Muchachos Observatory (ORM) on La Palma and Teide Observatory (TO) on Tenerife as these two sites were the candidates to host the European Solar Telescope (EST) \cite{quintero_noda2022}, the new generation solar telescope of 4.2 m in diameter.

A preliminary analysis has been performed with the Astro-Meso-Nh model above el Roque de los Muchachos and el Teide Observatories \cite{masciadri2024} by our team. From this paper we report here Fig.\ref{cn2_roque} showing the temporal evolution of the $C_N^2$ extended on the whole 20~km on the night and the contiguous day time on the date of 2009/08/17. Figure \ref{see_roque} shows on top the temporal evolution of the seeing on the nighttime and the contiguous daytime on the same date and on bottom the average of the $C_N^2$ in the whole night and day period. The convention used is that the date refers to the end of the simulation. For this reason, the date is the same for the nighttime and the daytime. On the x-axis we report the UT time (top) and the time elapsed from the starting time of the simulation (bottom).

 Looking at these two figures we can observe that the Astro-Meso-Nh model well reconstructs a bump of turbulence ($\CN2$) close to the surface in the middle of the daytime that is associated to the increasing value of the seeing in the central part of the daytime. This is due to the solar radiation of the daytime that produces a similar bump in Fig.\ref{cn2_roque} and Fig.\ref{see_roque} at the same time as expected. On the average of the $C_N^2$ profile we observe an important difference close to the surface as expected (due to the solar radiation) however we do not observe any specific difference between the nighttime and daytime in the free atmosphere. This is not surprisingly as there are no physical arguments justifying necessarily $\CN2$ differences at these heights between night and day. We highlight that the model used for these simulation is not calibrated for the optical turbulence. A not calibrated model does not allow a precise estimate in quantitative terms of the absolute value of astroclimatic parameters however it can provide very useful information in relative terms as relative estimates are consistent. 
 
 For this reason in Fig.\ref{see_20_comp} is shown the average value of the temporal evolution of the seeing above 20 dates (night and contiguous day) for the same dates above el Roque de los Muchachos and Teide Observatories. Above both sites we observe a peak of the seeing around midday and the height of the peak is comparable above the two sites. This lets us think that the two sites are characterised by comparable mean seeing. Our results are coherent with the decision taken on 2021 to select el Roque de los Muchachos at La Palma as the site hosting the EST telescope. 
 
 The fact that the model has not been calibrated is due to the fact that, at present time, the scientific community cannot access many reliable instruments conceived for the OT estimates during the day.  While for the nighttime we can count on a set of instruments that are commonly acknowledged by the scientific community as robust, development of instrumentation for OT estimates during the day is more recent and instrumentation is still in a phase of R\&D. This issue is however particularly relevant also for OT forecast and it is therefore a priority to boost this research activity. Among the instruments that appeared in the literature, that can cover the daytime and that we consider worth of attention we mention: (1) SHIMM24H that is a Shack-Hartmann Image Motion Monitor by Griffiths et al. 2023\cite{griffiths2023}, (2) SCO-SLIDAR that is a Single Coupled SLODAR and SCIDAR by Vedrenne et al. 2007\cite{vedrenne2007}, (3) a PDSL that is a Profiler for a Differential Solar Limb by Song et al.\cite{song2020} and (4) a S-DIMM that is a solar DIMM by Liu Zhong et al. 2000\cite{zhong2000}. The last instrument is not recent but it is, at our knowledge, the unique solar DIMM that is in operation at present time. The first three instruments are vertical profilers, the last one measures the integral turbulence along the 20~km.

\section{CONCLUSIONS}
\label{sec:concl}  % \label{} allows reference to this section

We summarize the main achievements obtained by our team in the context of optical turbulence forecast:\\  \\
\noindent
(1) We proved that is possible to forecast reliable $\CN2$ profiles and the most suitable tool are at present the atmospheric non-hydrostatic mesoscale models. Astro-Meso-Nh model has been the first model conceived for the optical turbulence using a non hydrostatic model providing $\CN2$ profiles. Such a model has been validated and applied to many among the best astronomical site of top-class telescopes in the world: Cerro Paranal, Cerro Armazones, Cerro Pachon and Cerro Pachon (Chile), Mt.Graham (Arizona, US), Roque de los Muchachos (La Palma, Canary Islands), Teide (Tenerife, Canary Islands), San Pedro Martir (Baja California, Mexico), Maidanak (Utzbekistan) and in remote placed on the Earth where is expected the minimum level of optical turbulence i.e. Dome C, Dome A, South Pole (Antarctica). These sites have been subject of great interest by the astronomical community in the [2000-2010] decade because of their excellent turbulence conditions and allowed us to achieve an exhaustive validation of the optical turbulence scheme. \\ \\
\noindent
(2) We provided the first calibration method for an optical turbulence parameterisation\cite{masciadri2001} for atmospheric mesoscale models that was of inspiration for other solutions for model calibration implemented later on.\\ \\
\noindent
(3) More recently we proposed the AR method, based on an autoregressive technique that is able to provide forecast performances never achieved before on a time scale of 1h and 2h that are the most relevant for the science operation of ground based astronomy. Forecast accuracy and performances for a FTS=1h are calculated using a rich sample of nights of one solar year. We obtained a RMSE = 0.1 arcsec for the seeing, 0.58 ms for the wavefront coherence time, 0.16 arcsec for the isoplanatic angle and 6\% for the ground later fraction (GLF). \\ \\
\noindent
(4) The gain of the forecast with the AR with respect to the standard configuration for a FTS =1h is of the order of [2.6, 3.8] depending on the astroclimatic and atmospheric parameters. We could also prove that the RMSE between AR forecast and observations for all all the astroclimatic parameters is smaller than the RMSE between the prediction by persistence and observations on a time scale from 1h to 4h\cite{masciadri2023}. The prediciton by persistence is taken as a method of reference and assumes that the value of a parameter remains constant for times larger than a time zero t$_{0}$. This achievement tells us that forecasts obtained with the AR method provide information that would not be accessible otherwise.\\ \\
\noindent
(5) Our team has been developer and it is responsible of two automatic operative forecast systems of the optical turbulence and atmospheric parameters relevant for ground-based astronomy above two among the most important top-class telescopes of present time in the world:\\ \\ 
\noindent
- {\bf ALTA Center at the Large Binocular Telescope (LBT)} \\ 
- {\bf FATE at the Very Large Telescope (VLT)} \\ \\
\noindent
Even if there are some differences between the two projects/tool, both of them deliver forecasts of the mentioned parameters at long and short time scales. More precisely:  (A) Long-time scales:  forecasts of all the parameters are delivered in the afternoon for the coming night. We call this standard configuration; (B) Short time scales:  forecasts at short time scales (1h or 2h are the most relevant ones) obtained with the AR method are delivered with an upgrade frequency of 10 minutes.  At our knowledge ALTA Center has been the first operation site in the world with such characteristics and performances. The AR method has been implemented first in 2019 with an upgrade of 1h, later on in 2024 with the up-grade frequency of 10 minutes.\\ \\
 \noindent
 (6)  Results published so far\cite{masciadri2023,turchi2022} indicate that machine learning (ML), deep learning, neural networking techniques did not overcome performances of the AR method and in some cases are even not better than the references methods (prediction by persistence or precasting). A deeper and more complete analysis with this approach is on-going and it will be possible to provide more precise information in this respect. We can anticipate that, due to the great flexibility of ML they can be applied to contexts where other techniques ca not play any role. They are therefore certainly worth of attention.   \\ \\
 \noindent
 (7) Experience gained so far tells us that we definitely entered in a new era in which hybrid techniques will play a fundamental role in the OT forecast at short forecast time regimes.   \\ \\
\noindent
(8)  We remained interested in improving the atmospheric models performances as long-term forecasts strongly depend on this approach. In other words, even though hybrid techniques certainly represent the future for short-term forecasts, the atmospheric models remain essential for long-term forecasts. \\ \\
\noindent
In terms of perspectives we mention a few main reflexions:\\  \\
\noindent
(1) We are entering into a new era in which hybrid techniques will play a fundamental role particularly for forecasts at short time scales (1h or 2h). With hybrid techniques we mean the numerical techniques i..e the atmospheric models plus statistical techniques (AR, machine learning, deep learning, neural networks, etc.)\\ \\
\noindent
(2) New challenges for the OT forecast applied to the free-space optical communication will certainly occupy investigations of scientific community in the coming years. FOS opens to a much larger field of applications than the ground-based astronomy. \\ \\
\noindent
(3) It is extremely important for both fields of application (ground-based astronomy and FOS) to boost the research activity in the field of instrumentation for OT estimates during the day time. These instruments are crucial and functional for an optimal exploitation of the OT forecasts and their potentialities.

\acknowledgments % equivalent to \section*{ACKNOWLEDGMENTS}       
 
The study is funded by the contract FATE N. PO102958/ESO/20/95952/FLAB and the contract ENV002 (LBTO). The authors acknowledge the LBT Director (Joe Shields), the ex-LBT Director (Christian Veillet), the LBT Board and the whole LBTO staff for supporting the ALTA Center. The authors acknowledge the responsible of the Science Operation Departement of VLT (Steffen Mieske), the ESO referent for the FATE project (Angel Otarola), Miska Le Louarn, the LaMMA institute in Florence that is a collaborator in the project FATE. We acknowledge the Tuscany Region for the contribution to the FATE project. The authors thanks the Meso-Nh user support team who constantly maintain the Meso-Nh model by developping new packages in successive model versions. The authors acknowledge the user support team of ECMWF.  Initialisation data of the Astro-Meso-Nh model come from the HRES model of ECMWF.

% References
\bibliography{report} % bibliography data in report.bib
\bibliographystyle{spiebib} % makes bibtex use spiebib.bst

\end{document}